%% file: 00-paper_top.tex
\documentclass[conference,letterpaper]{IEEEtran}
\usepackage{cite}
\usepackage{amsmath,amssymb,amsfonts}
\usepackage{graphicx}
\usepackage[dvipsnames,svgnames,table]{xcolor}
\usepackage[final]{microtype}
\usepackage[italic]{mathastext}
\usepackage[T1]{fontenc}
\usepackage{textcomp}
\usepackage[all]{nowidow}
\usepackage[keeplastbox]{flushend}
\usepackage{fancyhdr}

\pdfmapline{+hungergames <HungerGames.ttf <ec.enc}

\newcommand\hungergamestitlefont[1]{{\usefont{T1}{hungergames}{m}{n}\spaceskip=.30em #1}}

\usepackage[utf8]{inputenc}
\usepackage{tikz}
\usepackage{booktabs}
\usepackage{paralist}

\usepackage{enumitem}
\usepackage{comment}
\usepackage{subcaption}
\usepackage{tabularx}
\usepackage{makecell}
\usepackage{ragged2e}
\usepackage{threeparttable}
\usepackage[inkscapelatex=false]{svg}
\PassOptionsToPackage{hyphens}{url}
\usepackage{hyperref}
\usepackage{cleveref}
\crefname{algocf}{algorithm}{algorithms}
\Crefname{algocf}{Algorithm}{Algorithms}
\usepackage{caption}
\newcolumntype{L}{>{\RaggedRight\arraybackslash}X}
\newcolumntype{C}{>{\centering\arraybackslash}X}

\definecolor{tableheadspec}{HTML}{446688}

\hypersetup{
    colorlinks=true,
    linkcolor=ForestGreen,
    citecolor=ForestGreen,
    urlcolor=RoyalBlue,
    filecolor=IndianRed,
    pdftitle={PANEM: A Heuristic Latency Model}, 
    pdfauthor={Lucas Crowthers, Mahesh Madhav}, 
    pdfsubject={CPU Performance Modeling}, 
    pdfkeywords={PANEM, Ampere Computing, Memory Latency, Queueing, Workload Characterization, CPU Performance, Memory Simulation}, 
    pdfnewwindow=true,
    pdfdisplaydoctitle=true,
    bookmarksopen=true
}

\begin{document}
\bstctlcite{IEEEtran_bst_ctl}
\makeatletter
\newcommand\GoBig{\@setfontsize\Huge{25}{25}}
\makeatother

\title{\GoBig{\hungergamestitlefont{PANEM: A Heuristic Latency Model}}\\
\Large{Evaluating Core Behavior on Cloud Workloads}}
\author{

 \IEEEauthorblockN{Lucas Crowthers}
 \IEEEauthorblockA{\emph{Ampere Computing}}
 \IEEEauthorblockA{\emph{Raleigh, NC}}
 
 \and 
 \IEEEauthorblockN{Mahesh Madhav}
 \IEEEauthorblockA{\emph{Ampere Computing}}
 \IEEEauthorblockA{\emph{Portland, OR}}

}

\maketitle


\input{commands} 
\input{10-abstract}

\input{20-intro}
\input{30-background}
\input{40-methodology}

\input{50-results}
\input{60-limitations}
\input{80-conclusion}


\section*{Acknowledgments}

We used OpenAI Codex 5.4 for help with managing and plotting data; and Google Gemini 2.5 during the editing process for making word choices and refactoring prose into academic writing. All the text was reviewed and revised by the authors.
\clearpage
{\footnotesize
\bibliographystyle{IEEEtran}
\bibliography{90-citations}
}
\clearpage
\input{99-appendix}

\end{document}

%% file: commands.tex
\newcommand{\Red}[1]{{\color{red} #1}}
\newcommand{\ignore}[1]{}
\newcommand{\asm}[1]{\texttt{#1}}
\newcommand{\sys}[1]{\texttt{#1}}
\newcommand{\kw}[1]{\textit{#1}}
\newcommand{\kwb}[1]{\textbf{#1}}
\newcommand{\type}[1]{\textit{#1}}
\newcommand{\Response}[1]{{\color{blue} #1}}
\newcommand{\XXX}[1]{\Red{\textbf{XXX[}#1\textbf{]}}}
\newcommand{\tocite}[1]{\Red{CITE:\cite{#1}}}
\newcommand{\toref}[1]{\Red{REF:\ref{#1}}}
\newcommand{\parasub}[1]{\smallskip\noindent\textit{{#1:}\xspace}}
\newcommand{\mahesh}[1]{\textcolor{purple}{#1}}
\newcommand{\anyone}[1]{\textcolor{blue}{#1}}
\newcommand{\niparagraph}[1]{\noindent\textbf{\textsf{#1}\hspace{0.5em}}}
\newcommand\TODO[1]{\textcolor{red}{TODO: #1}}

\newenvironment{CompactItemize}%
  {\begin{list}{$\blacktriangleright$}%
    {\leftmargin=\parindent \itemsep=2pt \topsep=2pt
     \parsep=0pt \partopsep=0pt}}%
  {\end{list}}
\renewcommand{\labelitemi}{$\blacktriangleright$}

\newcommand{\malloc}{{\texttt{Malloc}}}
\newcommand{\linklist}{{\textsf{Linked-List}}}

\ExplSyntaxOn
\NewDocumentCommand{\anon}{m}
 {
  #1
 }
\ExplSyntaxOff


\definecolor{low}{HTML}{a31111} 
\definecolor{mid}{HTML}{FFFFFF}
\definecolor{high}{HTML}{a31111} 
\newcommand*{\opacity}{90}

\newcommand*{\minval}{0.500}
\newcommand*{\midval}{1.000}
\newcommand*{\maxval}{2.000}

\newcommand{\gradient}[1]{
    \ifdimcomp{#1pt}{>}{\maxval pt}{#1}{
        \ifdimcomp{#1pt}{<}{\minval pt}{#1}{
            \ifdimcomp{#1pt}{<}{\midval pt}{
                \pgfmathparse{int(round(100*(#1-\minval)/(\midval-\minval)))}
                \xdef\tempa{\pgfmathresult}
                \cellcolor{mid!\tempa!low!\opacity} #1
            }{
                \pgfmathparse{int(round(100*(#1-\midval)/(\maxval-\midval)))}
                \xdef\tempa{\pgfmathresult}
                \cellcolor{high!\tempa!mid!\opacity} #1
            }
            
    }}
}
\newcommand{\gradientbold}[1]{
    \ifdimcomp{#1pt}{>}{\maxval pt}{#1}{
        \ifdimcomp{#1pt}{<}{\minval pt}{#1}{
            \ifdimcomp{#1pt}{<}{\midval pt}{
                \pgfmathparse{int(round(100*(#1-\minval)/(\midval-\minval)))}
                \xdef\tempa{\pgfmathresult}
                \cellcolor{mid!\tempa!low!\opacity} \textbf{#1}
            }{
                \pgfmathparse{int(round(100*(#1-\midval)/(\maxval-\midval)))}
                \xdef\tempa{\pgfmathresult}
                \cellcolor{high!\tempa!mid!\opacity} \textbf{#1}
            }            
    }}
}

%% file: 10-abstract.tex
\begin{abstract}


Accurate pre-silicon memory modeling is essential for achieving meaningful representation of workloads on cloud-class many-core processors. Existing options force a poor tradeoff between fidelity and speed: fixed-latency models are fast but misleading, while cycle-accurate DRAM models are costly and difficult to scale across large study spaces or onto single-core environments. This paper presents PANEM, a lightweight event-driven heuristic model that has guided four generations of commercial server-core development at Ampere Computing. PANEM converts bandwidth-latency characterization data into a dynamic request-bytes/latency response, allowing miss latency to adapt to transient demand, queuing pressure, and read/write mix during simulation. Integrated into a single-core flow with configurable system-loading assumptions, PANEM enables realistic bandwidth constraints and contention-aware latency behavior without sacrificing throughput. Across a broad cloud workload trace suite, PANEM avoids the optimistic and pessimistic biases of fixed-latency baselines, yields more reliable conclusions for prefetching and dynamic throttling studies, and materially improves core-resource sizing decisions. These results show that a calibrated, contention-aware abstraction can deliver practical predictive value for industrial design-space exploration at simulation costs similar to fixed-latency models.
\end{abstract}

%% file: 20-intro.tex
\section{Introduction}




The performance of modern microprocessors is inextricably linked to the efficiency of the memory subsystem. In the era of many-core Systems-on-Chip (SoCs) prevalent in datacenter environments, where numerous cores compete for shared resources like caches and memory bandwidth, accurately predicting and optimizing for memory behavior is no longer a secondary concern but a primary design pillar. The latency experienced on a cache miss is not a static value; it is a dynamic outcome of system-wide contention. Consequently, pre-silicon performance simulation, a cornerstone of microprocessor design, must accurately capture this complex interplay to be of predictive value.

The computer architecture community has rightly focused on this challenge, striving to bridge the gap between simulation and silicon reality. Recent studies highlight the persistent difficulties in achieving high-fidelity memory simulation. Research has quantified these discrepancies, pinpointing the CPU-memory interface as a primary source of model inaccuracy \cite{diff_perspectives} and identifying missing latency as a cause of miscorrelation with hardware \cite{missing_link}. Workload behavior can also interact with DRAM organization and policy in complex ways \cite{ghose_dram_interactions}. While cycle-accurate DRAM simulators, including Ramulator \cite{ramulator} and DRAMSim3 \cite{dramsim3}, can offer high fidelity, their runtime cost in system-level simulations \cite{zsim_paper, rethinking_dram_sim} and significant configuration complexity \cite{mess_debunk} can make large design-space sweeps impractical. In response, high-level architectural simulation \cite{interval_simulation} and alternative memory mechanisms, from event-driven models \cite{event_based_dram_model} to immediate-response abstractions \cite{immediate_response}, seek a better balance of speed, adaptability, and accuracy \cite{rethinking_dram_sim}. Parallel to these tooling advancements, a consensus is emerging on evaluation philosophy; namely, that new core microarchitecture features must be evaluated in constrained-bandwidth environments to stress-test their effectiveness in realistic system conditions \cite{dpc4_summary, dpc4_talks, limoncello, themis}.


While these explorations provide invaluable insight, the challenge for architects of commercial cloud server products is fundamentally predictive. The unique demands placed on our products require performance to be evaluated across a vast matrix of different workloads, system loading levels, and memory configurations. A lack of clear expectations for memory bandwidth and latency early in the design process coupled with a desire to achieve study throughput necessary to answer these questions could result in a choice of fast but naive models for extra-core behavior such as fixed miss latency. But a new core cannot be designed in a vacuum; its performance characteristics, such as internal buffer sizing and prefetcher efficacy, are deeply influenced by the backpressure exerted by a congested system. Designing a core with an overly simplistic memory model results in choices that are incorrectly sized and will fail to meet performance targets in a real, shipping product. The point for industry practitioners is to think beyond making a simulator match existing hardware, to creating a highly configurable system-level representation that accurately models the target SoC ecosystem for hardware that does not yet exist.


This paper presents PANEM, a lightweight memory modeling methodology successfully used to design four generations of the AmpereOne\textregistered ~server CPU core \cite{ampereone_perf, AmpereOneX_Brief, AmpereOneM_Brief, ampere_mte}. This work makes the following key contributions: First, we detail a novel, heuristic technique that transforms empirical bandwidth-latency curves into a request-bytes/latency lookup table to dynamically determine latency based on system load. Second, we demonstrate that this method accurately models system behavior beyond the saturation point, a key limitation of prior analytical models. Finally, we provide an analysis showing that our model offers superior predictive capability for microarchitectural design decisions compared to fixed-latency models, enabling more efficient and accurate pre-silicon design of high-performance cores for many-core SoCs.

%% file: 30-background.tex
\section{Background}
Panthera is the name for \anon{Ampere}'s simulation environment and includes detailed core models, mesh models, various memory models, and tooling for workload trace capture, study execution, and data analysis \cite{ampere_pv, ampere_mav}. We focus on the environment within Panthera for detailed modeling of a single core and how we achieve a meaningful level of accuracy for the representation of system-level effects that such a core deployed in a real cloud workload environment would encounter, particularly the share of system resources allotted to it while competing with other (unsimulated) cores executing workloads at the same time.

\begin{figure}[ht] 
    \centering
    \includegraphics[width=\columnwidth]{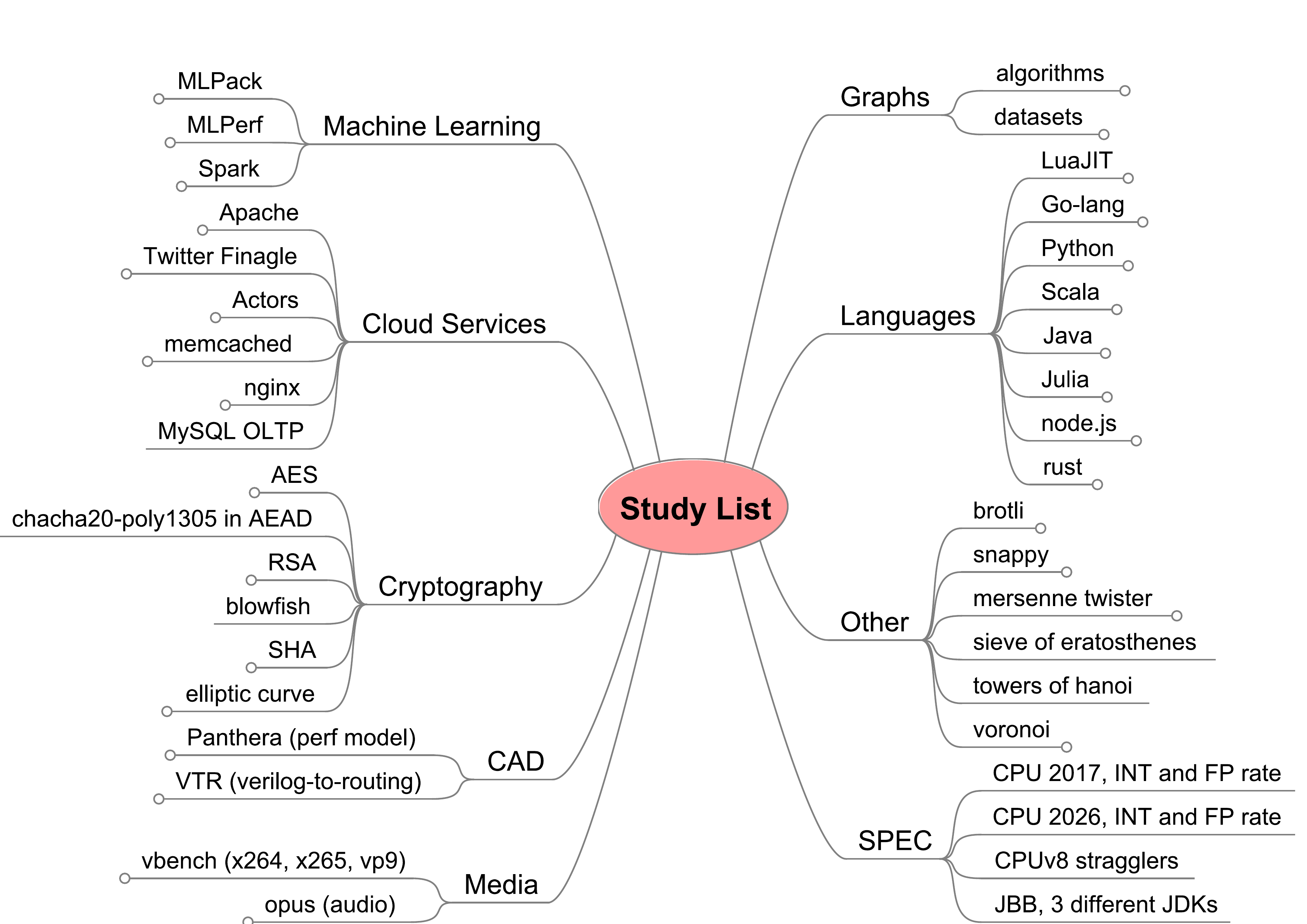}
    \caption{Sample taxonomy of workloads in the core study list.}
    \label{fig:study list}
\end{figure}

\anon{Ampere}'s typical core study list consists of about 4300 traces from 300 distinct workloads. These traces are usually 10M instructions in length with a workload-appropriate level of warmup for both memory and branch predictor state. A high-level breakdown of workloads from which these traces are derived is depicted in \autoref{fig:study list}.

Cloud service providers use vastly different system loading scenarios for different levels of service and different workloads. While it may be obvious that performance under heavily loaded scenarios is important, competitive cloud servers must also demonstrate performance at single- and low-core counts to be successfully considered for deployment. As such, any evaluation of architectural performance features must consider the impacts on power and performance across a vast space of possible configurations. This places demands on both engineering time to set up and analyze the results of such experiments, and on the raw simulation time to complete them. In order to maximize efficiency of this work, \anon{Ampere} not only simulates full system level behavior but also strives to build meaningful representations of these system-level effects into single core simulation.

For such a configuration within Panthera, the detailed core model--which models out to the L2 cache's external request interface--is connected to a simplified but representative mesh supporting hash-based request routing and a distributed last-level cache. To represent different loading configurations, this LLC is fractionally reduced or increased to give an appropriate per-core share of cache. Misses at this LLC are routed to an array of memory controllers which can be populated with a variety of models available within Panthera. Although this configuration leads to more realistic LLC and heterogeneous idle latency behavior due to different routing path lengths \cite{missing_link}, a side effect is significant over-provision of mesh tracking resources, throughput, and memory capability for a single core.

Because of this, the use of a cycle-accurate DRAM simulator requires the synthesis of a meaningful level of bandwidth demand. This either calls for multi-core simulation or some sort of traffic replication scheme, either of which entails more configuration and runtime overhead in addition to that already introduced via costly DRAM models. Thus, to achieve the goal of creating a representative environment for single-core simulation while maintaining a performance level necessary to address product goals, a unique solution is required: PANEM (Panthera's New Memory Model).

%% file: 40-methodology.tex
\section{Methodology}
PANEM is a lightweight, event-driven heuristic memory model. It is designed to provide a realistic, dynamic latency response based on system load, using standard bandwidth/latency curves as its primary input. The methodology can be broken down into two phases: an offline preparation step and a runtime operational loop.

\subsection{Offline Curve Transformation}

PANEM begins with a standard bandwidth/latency curve, which characterizes a memory system's response to increasing bandwidth demand and can be sourced from hardware measurements or vendor data. Using Little's Law \cite{little}, this bandwidth/latency data is converted into a \texttt{request-bytes/latency} lookup table. This transformation is crucial: while a bandwidth/latency curve is not a proper function beyond the saturation point (i.e., bandwidth may drop as latency continues to increase), the resulting \texttt{request-bytes/latency} relation is. This allows PANEM to model system behavior gracefully even under heavy oversaturation, a key advantage over analytical models like ZSim's M/D/1, which must clamp utilization below 100\% to avoid its formula's asymptote. The model can be parameterized with multiple curves to represent different read/write ratios.


\subsection{Runtime Latency Calculation}

During simulation, PANEM operates on an event-driven loop:


\begin{enumerate}
    \item \textbf{Demand Tracking:} The model tracks all in-flight memory requests. Upon arrival, read and write requests are placed into separate queues. While write requests can be acknowledged immediately to the core, they remain tracked internally to contribute to system load. Over a configurable sliding time window (a single tumbling 520\,ns window in our default configuration), PANEM calculates the time-weighted average of outstanding requests and the read/write ratio.

    \item \textbf{Latency Update:} Periodically, a timer triggers a latency update. Using the calculated average outstanding requests and read/write ratio from the preceding window(s), PANEM selects the appropriate pre-computed lookup table and determines a new \textit{predetermined latency}. Latency values are selected by interpolating linearly between the nearest points on the selected curve.

    \item \textbf{Request Servicing:} All new requests arriving at the memory model are assigned a completion time based on the current \textit{predetermined latency}. When a request's latency timer expires, it is dequeued and a response is sent if the request was a read.
\end{enumerate}

\noindent This mechanism creates a feedback loop where a higher number of outstanding requests leads to a higher lookup latency in the next window, naturally modeling the contention and queuing effects of a loaded memory subsystem. This approach allows PANEM to accurately capture the performance impact of system-level effects within a fast, single-core simulation environment. See \autoref{fig:flowchart} in the appendix for a flowchart.


\subsection{Comparison and Design Rationale}


Although the Mess simulator \cite{mess} is also parameterized via a family of bandwidth/latency curves, it uses an iterative process to progressively approximate the correct latency for a given workload phase. On the other hand, PANEM can directly select an appropriate latency based on recently observed workload behavior. Thus, PANEM can achieve a high degree of accuracy in steady-state while also being more reactive to transient workload behavior than an iterative simulator like Mess. Consequently, PANEM's design is best understood in comparison to analytical queuing models like ZSim's M/D/1 implementation. Both approaches aim to model contention by updating a response latency based on measured demand. However, PANEM makes several design choices that prioritize empirical accuracy and the ability to model system oversaturation, as summarized in \autoref{tab:panem_vs_md1}.

\begin{table}[!ht]
    \centering
    \begin{tabular}{p{0.45\columnwidth} p{0.45\columnwidth}}
        \textbf{M/D/1} & \textbf{PANEM} \\
        \hline
        Averages incoming request rate over time & Time-weighted average of outstanding requests \\
        \hline
        Relates latency and utilization \cite{zsim_md1} via Pollaczek-Khinchine formula \cite{pollaczek, khinchine} & Prepares a lookup table from a provided bandwidth/latency curve \\
        \hline
        Must clamp utilization to $<$100\% due to asymptote in Pollaczek-Khinchine & Can react to over-utilization if supplied with an appropriate input curve \\
    \end{tabular}
    \caption{Key differences between ZSim's M/D/1 queuing and PANEM.}
    \label{tab:panem_vs_md1}
\end{table}

\begin{figure}[h] 
    \centering
    \includegraphics[width=\columnwidth]{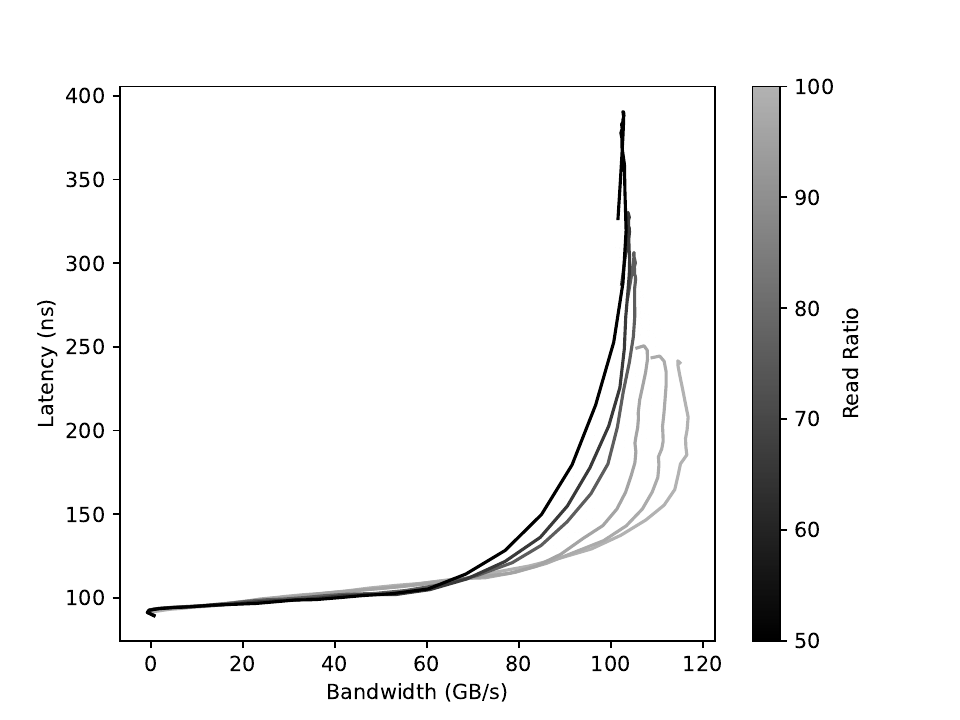}
    \caption{A representative subset of the 26 bandwidth/latency curves reported in the Mess paper \cite{mess}.}
    \label{fig:mess_skylake_bw_reconstructed}
\end{figure}

\begin{figure}[h] 
    \centering
    \includegraphics[width=\columnwidth]{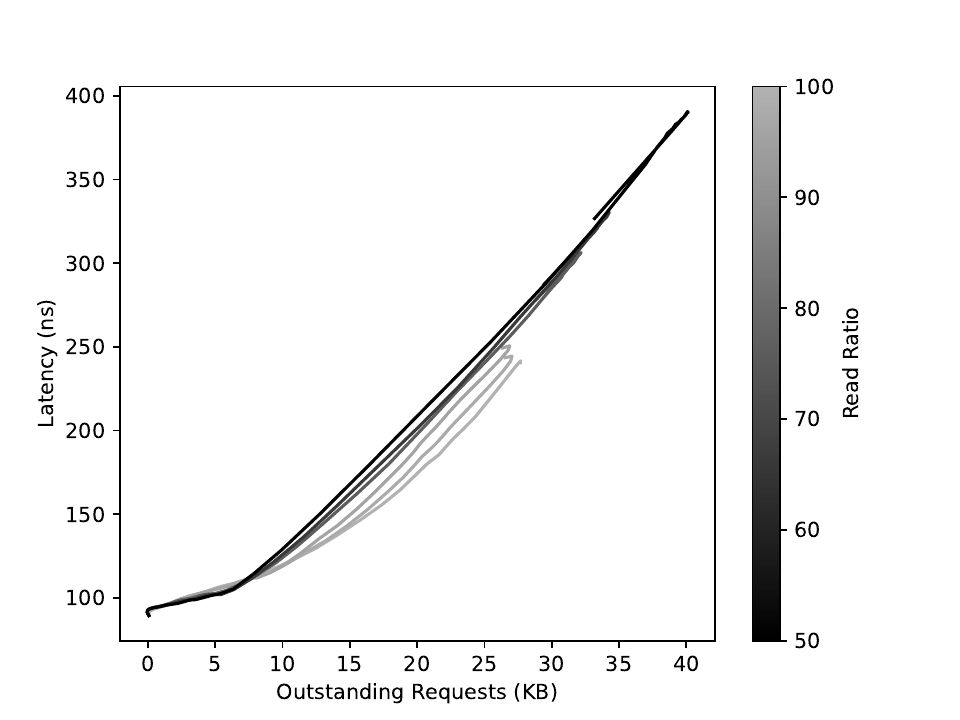}
    \caption{Example conversion of curves depicted in \autoref{fig:mess_skylake_bw_reconstructed} into corresponding request-bytes/latency representation.}
    \label{fig:mess_skylake_req_bytes}
\end{figure}

\begin{figure}[h] 
    \centering
    \includegraphics[width=\columnwidth]{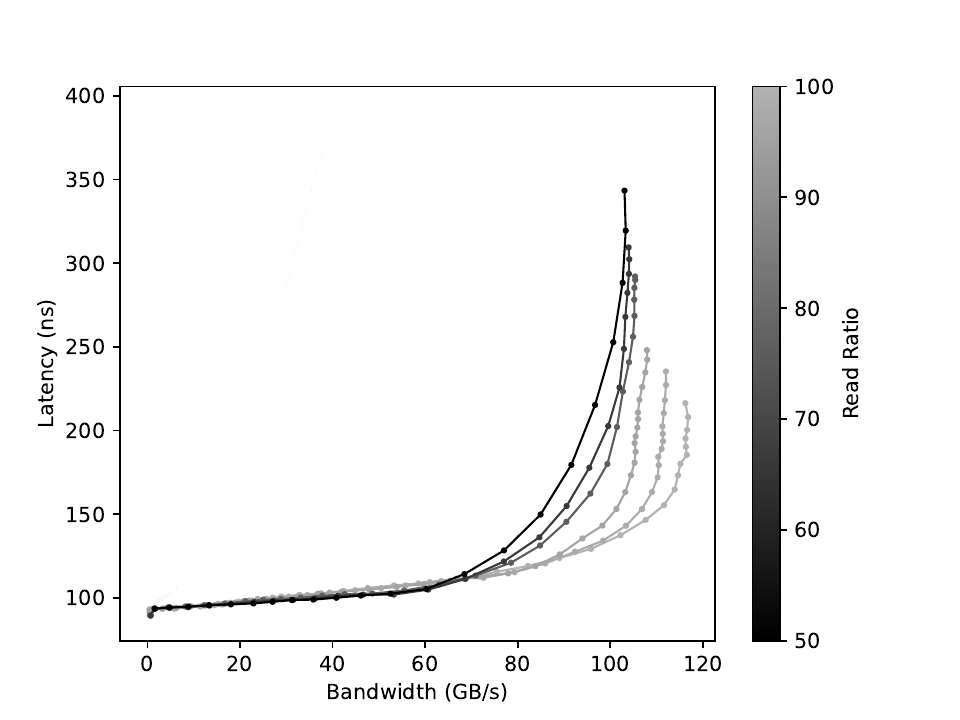}
    \caption{Filtered bandwidth/latency curves from \cite{mess} used as input to PANEM.}
    \label{fig:mess_skylake_bw_filtered}
\end{figure}

\begin{figure}[h] 
    \centering
     \includegraphics[width=\columnwidth]{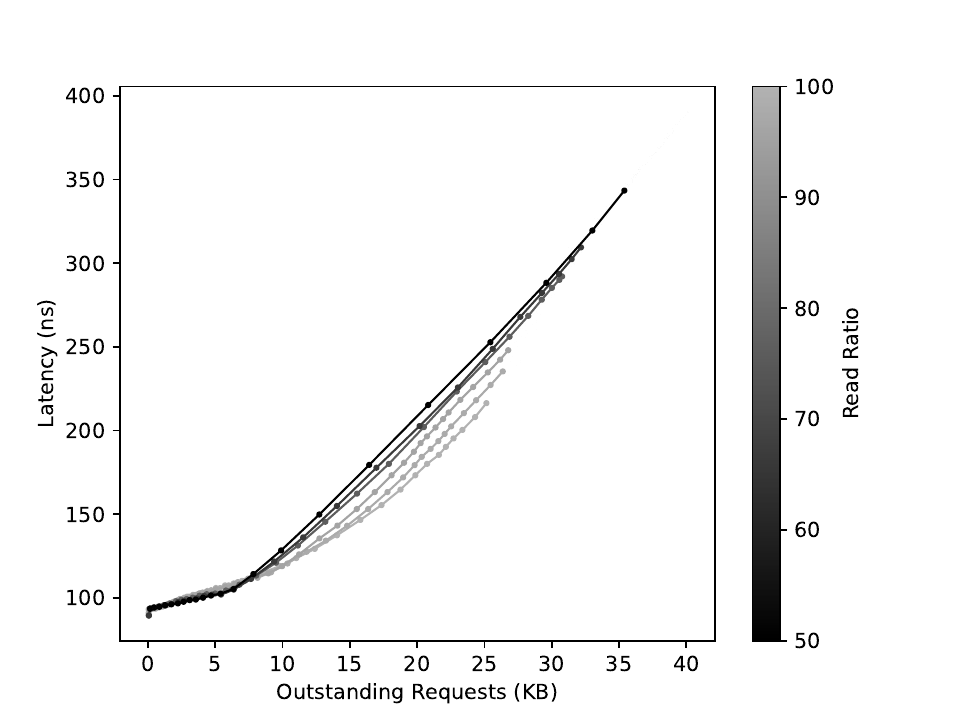}
    \caption{Request-bytes/latency representation of \autoref{fig:mess_skylake_bw_filtered}.}
    \label{fig:mess_skylake_req_filtered}
\end{figure}

The most significant distinction lies in how PANEM handles system saturation. By converting empirical bandwidth/latency data into a \texttt{request-bytes/latency} relation, PANEM creates a lookup model that remains well-defined even when demand exceeds peak bandwidth. To illustrate the simplicity of configuration and achievable accuracy, we ingest previously reported Intel Skylake bandwidth/latency curves \cite{mess}, a subset of which are plotted in \autoref{fig:mess_skylake_bw_reconstructed}. The converted request-bytes/latency representation of this data is depicted in \autoref{fig:mess_skylake_req_bytes}. Both of these curve families show sharp reflections near the bandwidth/latency envelope which we presume to be noise introduced by the measurement methodology rather than meaningful characterization of the system's behavior. Additionally, the request-bytes/latency curves show a stabilization of behavior even before the bandwidth peak while the bandwidth/latency curves are still turning over. In order to smooth out the curves we remove any points in the request-bytes/latency relation that do not increase both outstanding request-bytes and latency relative to the previous point and remove all but the first point after the bandwidth peak is achieved in each curve. This gives the converted bandwidth/latency relation ready for direct ingestion into PANEM without any further adjustment (\cref{fig:mess_skylake_bw_filtered,fig:mess_skylake_req_filtered}). Although these figures show only a small subset of the 26 individual curves reported in the Mess paper \cite{mess}, the full family of curves were converted and ingested into PANEM. The transformed curve continues to map higher outstanding request bytes to higher latencies. PANEM presumes this trend continues to arbitrarily high request-byte values by extrapolating linearly from the last two points in each curve. This is critical for modeling the performance collapse observed in real hardware, where bandwidth decreases as the system becomes overly congested.


This data-driven parameterization also provides a significant practical benefit. Because bandwidth/latency curves are often the earliest meaningful characterization data available from IP vendors that describe memory and memory controller behavior, PANEM can be quickly configured to model future DRAM technologies, accelerating our core and system co-design process.

\begin{figure}[h] 
    \centering
    \includegraphics[width=\columnwidth]{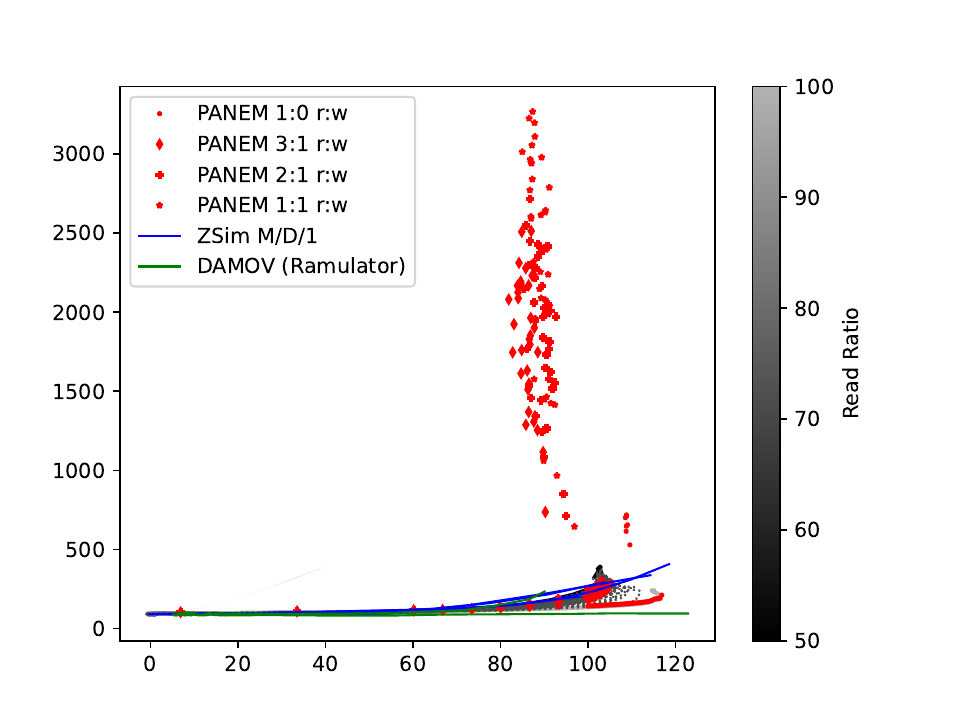}
    \caption{Response of PANEM when driven by an abstract traffic generator and configured to reproduce ingested bandwidth/latency curve data captured from Intel Skylake \cite{mess}.}
    \label{fig:panem_skylake_response}
\end{figure}

\begin{figure}[h] 
    \centering
    \includegraphics[width=\columnwidth]{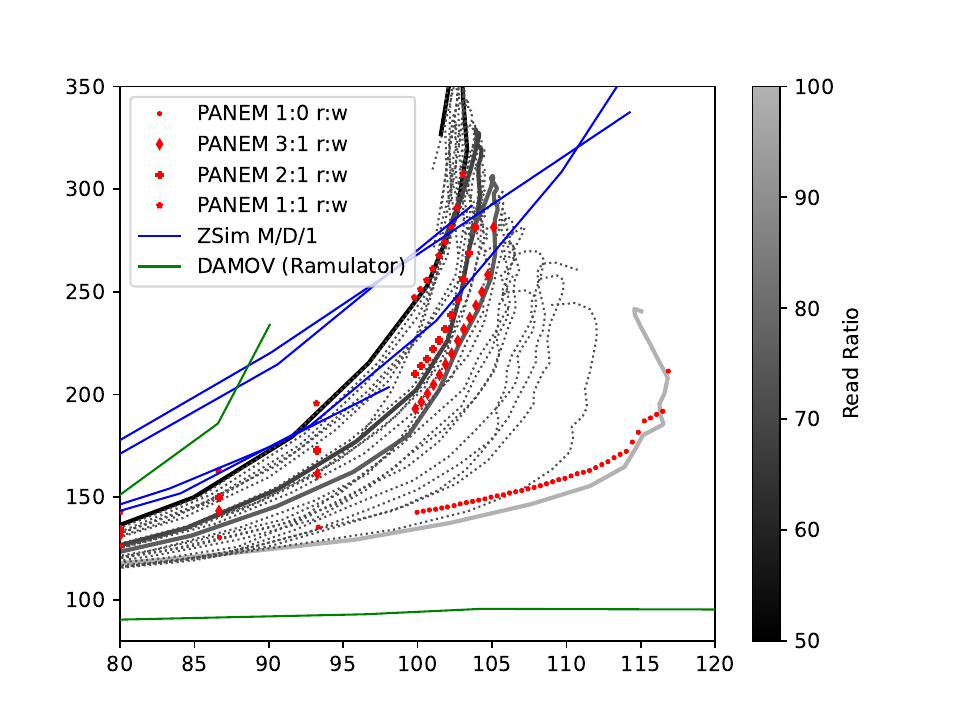}
    \caption{Finer detail of PANEM response when driven by abstract traffic generator. Original bandwidth/latency curves corresponding to the nominal read:write ratios of the 4 workloads are accentuated.}
    \label{fig:panem_skylake_response_detail}
\end{figure}

In order to evaluate the accuracy of PANEM configured with these converted bandwidth/latency curves, we attached two abstract cores to a router and a single PANEM model. PANEM was configured with a JSON representation of the filtered data extracted from original curves. The router and cores were sized to be capable of delivering 256 GB/s of bandwidth to ensure that they do not bottleneck the 128 GB/s theoretical peak of the DDR4-2666 memory characterized within the curves. The cores are able to execute a small subset of Arm instructions and track dependencies between them. One core executes a pseudo-random pointer-chase to measure load-use latency while the other executes one of a number of bandwidth-generation workloads. Each core is throttled by a request credit return rate limiter which allows us to sweep bandwidth of the traffic generation core. By sweeping bandwidth and measuring latency while executing 4 workloads consisting of independent load and store instructions interleaved in different ratios (1:0 read:write, 3:1 read:write, 2:1 read:write, and 1:1 read:write) we can measure the bandwidth/latency response of PANEM within this system and compare it to the ingested curve response. The results of this sweep are shown in \autoref{fig:panem_skylake_response} and \autoref{fig:panem_skylake_response_detail}. We also plot the ZSim M/D/1 results from \cite{mess} and the DAMOV (Ramulator) results from \cite{mess_debunk} which are intended to represent this same system. PANEM much more closely replicates the target behavior than either of these models, especially near the bandwidth peak. Since the abstract cores are only rate-limited and functionally unlimited in number of outstanding requests compared to the memory capacity, any demanded request rate above the peak experiences progressively degraded performance. This behavior would be muted in a more realistic system with greater constraints on total outstanding requests. Increased variance at very high demand rates is due to retry traffic from the memory configuration becoming fully saturated. The curves were not adjusted beyond the filtering process described above and reported latency for each workload is total load-use at the abstract core. The important takeaway is that even for unrealistically high sustained request rates PANEM constrains bandwidth to below the peak while imposing increasing latency. The latency offset in PANEM along each curve is due to the time associated with router transit and execution of instructions by the abstract core. When deployed in more realistic models PANEM provides a parameter to subtract the effect of modeled behavior between PANEM and the core to give the expected idle load-use latency.

%% file: 50-results.tex
\section{Results}

\subsection{Hardware Correlation Comparison}

\autoref{tab:cpu2017_results} breaks down model projections for SPEC CPU2017 benchmarks \cite{cpu2017} relative to measured results for a half and fully-loaded real-world hardware system with many cores. This data was originally presented by our colleagues to describe their work on Memory Access Vectors for workload characterization \cite{ampere_mav}. PANEM was crucial in the correlation of these results by affording the ability to run reliable bandwidth sensitivity sweeps on the core model. We compare their results against two fixed-latency configurations: a low bound equal to the idle latency of our curve and a high bound equal to the latency experienced at peak read bandwidth. These latency values represent likely options for evaluating core performance if limited to fixed-latency memory models. PANEM achieves better overall correlation than either fixed configuration, but more importantly maintains much closer accuracy and consistency at the individual benchmark level. This is especially true for the fully loaded configuration where idle fixed latency drastically overestimates performance and high fixed latency performs better mostly by over- and under-estimating individual benchmark performance to more equal degrees. PANEM's ability to provide accurate results at benchmark granularity helps ensure that evaluated features are neither elected nor rejected based on unrealistic swings in performance due to drastic interactions between single benchmarks and the memory response model. Additionally, while 523.xalancbmk improves for all configurations under MAV characterization, the improvement in PANEM runs is much greater than for fixed configurations. For certain workloads like 523.xalancbmk, meaningful characterization is only possible within a sufficiently representative system environment.

\begin{table*}[!ht]
    \centering
    \caption{Performance Correlation for SPEC CPU2017 Integer Benchmarks \cite{cpu2017}, between memory models and final silicon running multicopy `refrate' \cite{ampere_mav}. The goal is to correlate at 1.0, which means the model provides score projections exactly matching silicon. The data shows the PANEM-based model correlates much closer to the many-core SoC under test compared to the fixed latency models.}
    \label{tab:cpu2017_results}
    \begin{tabular}{l ccc c ccc}
        \toprule
        & \multicolumn{3}{c}{\textbf{Half Loaded System (96 cores)}} & & \multicolumn{3}{c}{\textbf{Fully Loaded System (192 cores)}} \\
        \cmidrule(lr){2-4} \cmidrule(lr){6-8}
        \textbf{Benchmark} & PANEM & Idle Fixed & High Fixed & & PANEM & Idle Fixed & High Fixed \\
        \midrule
        500.perlbench\_r & \gradient{0.99} & \gradient{0.99} & \gradient{0.82} & & \gradient{0.98} & \gradient{1.03} & \gradient{0.83} \\
        502.gcc\_r       & \gradient{1.06} & \gradient{1.22} & \gradient{0.86} & & \gradient{1.05} & \gradient{1.50} & \gradient{1.04} \\
        505.mcf\_r       & \gradient{0.88} & \gradient{1.17} & \gradient{0.78} & & \gradient{1.03} & \gradient{1.95} & \gradient{1.28} \\
        520.omnetpp\_r   & \gradient{1.04} & \gradient{1.18} & \gradient{0.70} & & \gradient{1.01} & \gradient{1.42} & \gradient{0.82} \\
        525.x264\_r      & \gradient{0.99} & \gradient{1.00} & \gradient{0.97} & & \gradient{0.99} & \gradient{1.04} & \gradient{1.00} \\
        531.deepsjeng\_r & \gradient{1.06} & \gradient{1.06} & \gradient{0.92} & & \gradient{1.08} & \gradient{1.09} & \gradient{0.94} \\
        541.leela\_r     & \gradient{0.99} & \gradient{1.00} & \gradient{0.99} & & \gradient{0.97} & \gradient{1.00} & \gradient{0.99} \\
        548.exchange2\_r & \gradient{1.02} & \gradient{1.02} & \gradient{1.01} & & \gradient{1.02} & \gradient{1.02} & \gradient{1.02} \\
        557.xz\_r        & \gradient{0.91} & \gradient{0.93} & \gradient{0.64} & & \gradient{0.93} & \gradient{0.99} & \gradient{0.68} \\
        \midrule
        523.xalancbmk\_r (base) & \gradient{0.84} & \gradient{0.95} & \gradient{0.70} & & \gradient{0.80} & \gradient{1.15} & \gradient{0.84} \\
        523.xalancbmk\_r (MAV) & \gradient{0.95} & \gradient{1.02} & \gradient{0.75} & & \gradient{0.98} & \gradient{1.19} & \gradient{0.89} \\
        \midrule
        \textbf{geomean (with MAV)} & \gradientbold{0.99} & \gradientbold{1.06} & \gradientbold{0.84} & & \gradientbold{1.00} & \gradientbold{1.19} & \gradientbold{0.94} \\
        \bottomrule
    \end{tabular}
\end{table*}

\subsection{Bandwidth/Latency Distributions}

\begin{figure}[!h] 
    \centering
    \includegraphics[width=\columnwidth]{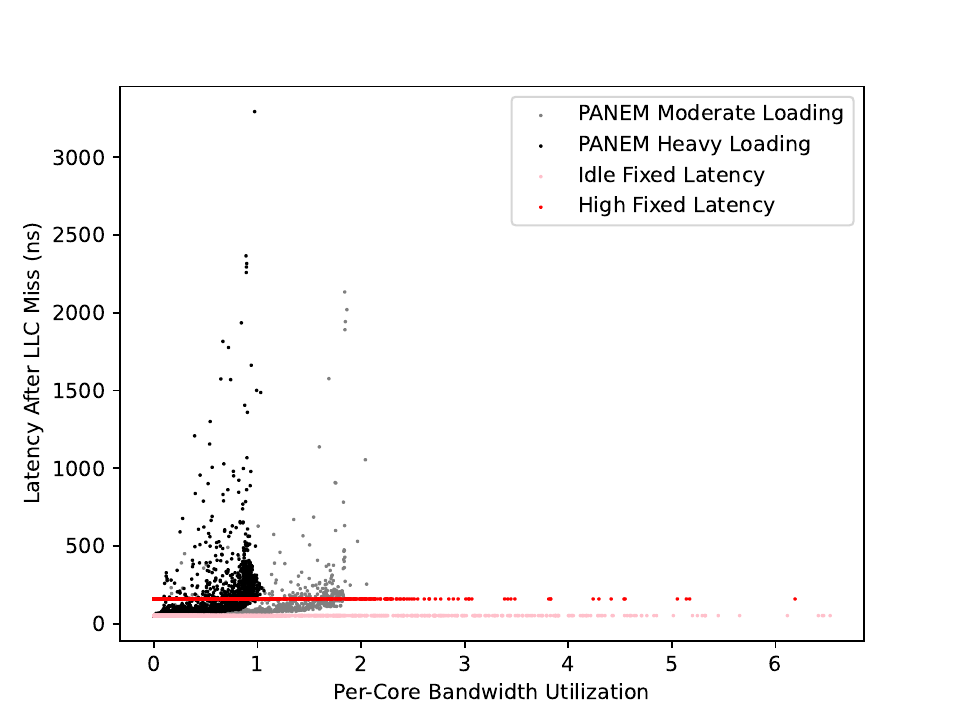}
    \caption{Comparison of bandwidth and latency distribution between PANEM and fixed-latency memory models. Traces in both fixed-latency configurations consume a much wider variety of bandwidth than PANEM.}
    \label{fig:panem_fixed_latency_comparison}
\end{figure}

\begin{figure}[ht] 
    \centering
    \includegraphics[width=\columnwidth]{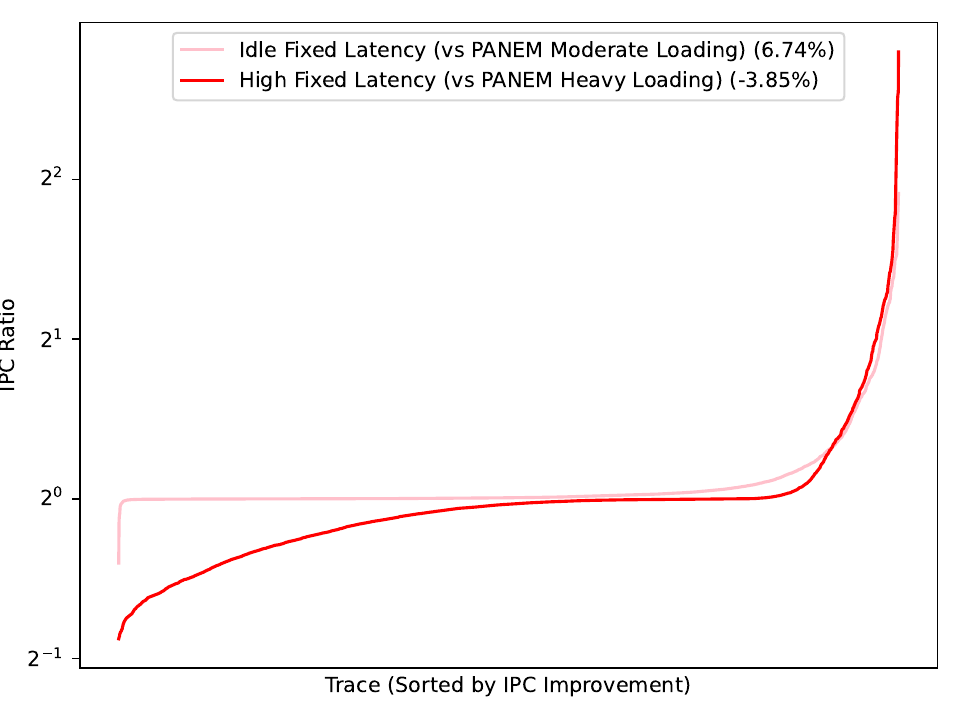}
    \caption{Comparison of IPC impact between PANEM and fixed-latency memory models. Both fixed-latency configurations are highly optimistic compared to PANEM but the idle fixed-latency configuration is simultaneously pessimistic for a wide portion of the study list.}
    \label{fig:panem_vs_fixed_ucurves}
\end{figure}

\autoref{fig:panem_fixed_latency_comparison} demonstrates the bandwidth-constraining capabilities of PANEM when compared to fixed response latency models. Bandwidth is normalized to the per-core share for a fully populated and utilized system. Each point in the scatter plots represents the average bandwidth and latency for one trace within our study list. All four of these configurations were run with L2 prefetchers and reactive bandwidth throttling (Completer Busy \cite{arm_cbusy}) disabled at the core to accentuate the impact of memory model choice. Even when a relatively high latency is used for all requests, the fixed-latency models frequently deliver drastically higher average bandwidths than PANEM. While the idle latency fixed configuration significantly overpredicts performance relative to PANEM moderate loading, the high fixed latency configuration underpredicts performance in spite of this high achieved bandwidth even when compared to PANEM heavy loading (\autoref{fig:panem_vs_fixed_ucurves}).

\begin{figure}[ht] 
    \centering
    \includegraphics[width=\columnwidth]{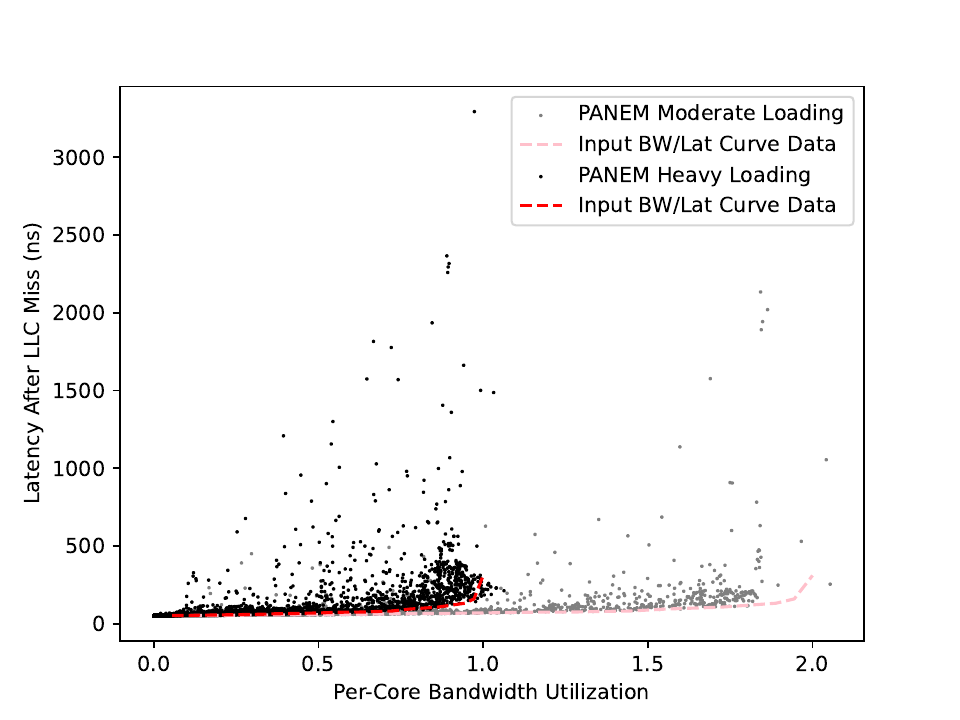}
    \caption{Detail of different bandwidth constraints defined by PANEM configuration. Distinct core behavior exposed by these configurations is already evident in the differing distributions near peak bandwidth in each case.}
    \label{fig:panem_loading_configs}
\end{figure}

On the other hand, both PANEM configurations cleanly constrain bandwidth to their respective curve parameters as demonstrated in \autoref{fig:panem_loading_configs}. PANEM also demonstrates a loss of bandwidth and increasing latency at higher request rates, similar to that observed in real hardware \cite{mess}. For these two scenarios the same base input bandwidth/latency curve is used and bandwidth is scaled with a single model parameter to achieve the new constraint, reducing the chance of misconfiguration. Compared to the arbitrary bandwidths achieved with fixed-latency models, PANEM much more closely conforms to the characteristic reported for actual hardware.

\subsection{L2 Prefetcher Impact}

\begin{figure}[ht] 
    \centering
    \includegraphics[width=\columnwidth]{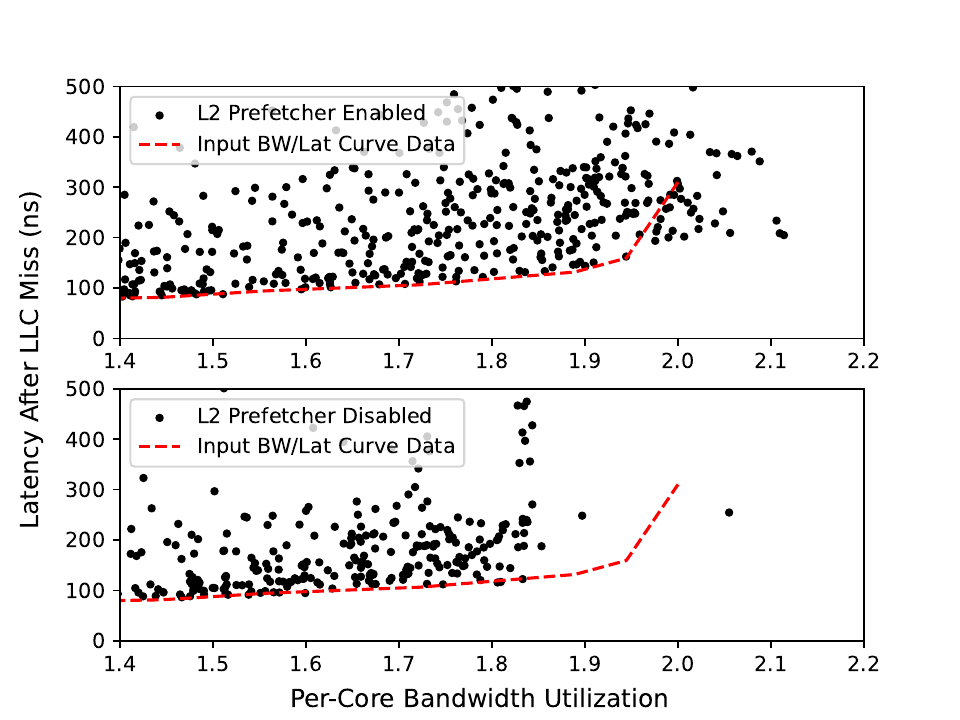}
    \caption{Comparison of workload response to PANEM bandwidth constraints with and without L2 prefetching. Unlike the fixed-latency model, the greater bandwidth available in this configuration is not fully utilized without prefetching.}
    \label{fig:panem_prefetcher_comparison}
\end{figure}

Notice that while the single core can overrun expected bandwidth limitations in both fixed-latency configurations and saturate bandwidth in many traces for the PANEM Heavy Loading configuration, almost no traces saturate the PANEM Moderate Loading configuration. By re-enabling L2 prefetchers we can recover this lost bandwidth (\autoref{fig:panem_prefetcher_comparison}) resulting in an increase in average bandwidth of 22\% while latency increases by 10\%. A number of traces can be seen to exceed the defined input curve represented by the red dashed line. This is a predictable phenomenon of the averaging window and how latency is applied, and their cause and impact are discussed in greater detail later in \autoref{sec:limitations}.

\begin{figure}[ht] 
    \centering
    \includegraphics[width=\columnwidth]{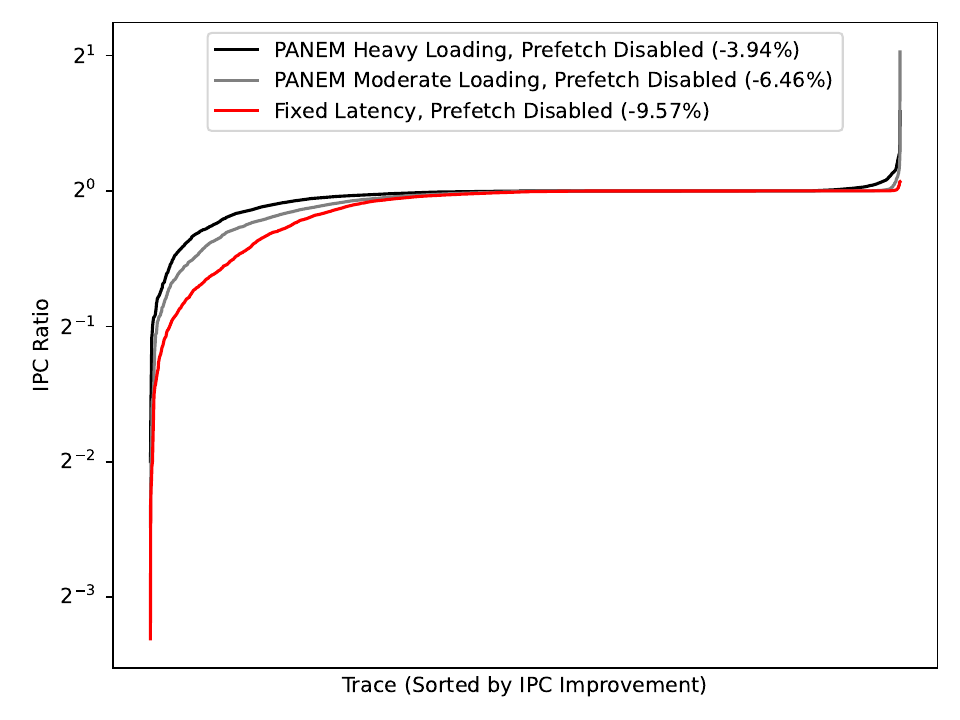}
    \caption{IPC impact of disabling L2 prefetchers. Greater sensitivity in fixed models is due to the loss of broad cases of marginally helpful prefetches. Under PANEM the loss of these prefetches does increase latency but also makes more bandwidth available to demand loads, a tradeoff that reduces the impact of their loss.}
    \label{fig:no-prefetch-ucurve}
\end{figure}

Considering the impact of prefetchers, PANEM gives more insight than a fixed-latency model (\autoref{fig:no-prefetch-ucurve}). We might expect that the low latency of this model--which is always lower than any response in either PANEM configuration--would lead to a reduced sensitivity to prefetcher behavior. However, \autoref{fig:no-prefetch-ucurve} shows the opposite: sensitivity is highest for the fixed-latency model (48\% higher than PANEM moderate) and lowest for PANEM heavy, the configuration with the highest average latencies. While this seems counterintuitive, the fact that successful prefetches are more valuable in the higher latency PANEM configurations is offset by the fact that unhelpful prefetches serve to consume available bandwidth. In other words, under fixed-latency models, prefetches only compete with demand loads for cache capacity, while under PANEM they compete for both cache and bandwidth.

\begin{figure}[ht] 
    \centering
    \includegraphics[width=\columnwidth]{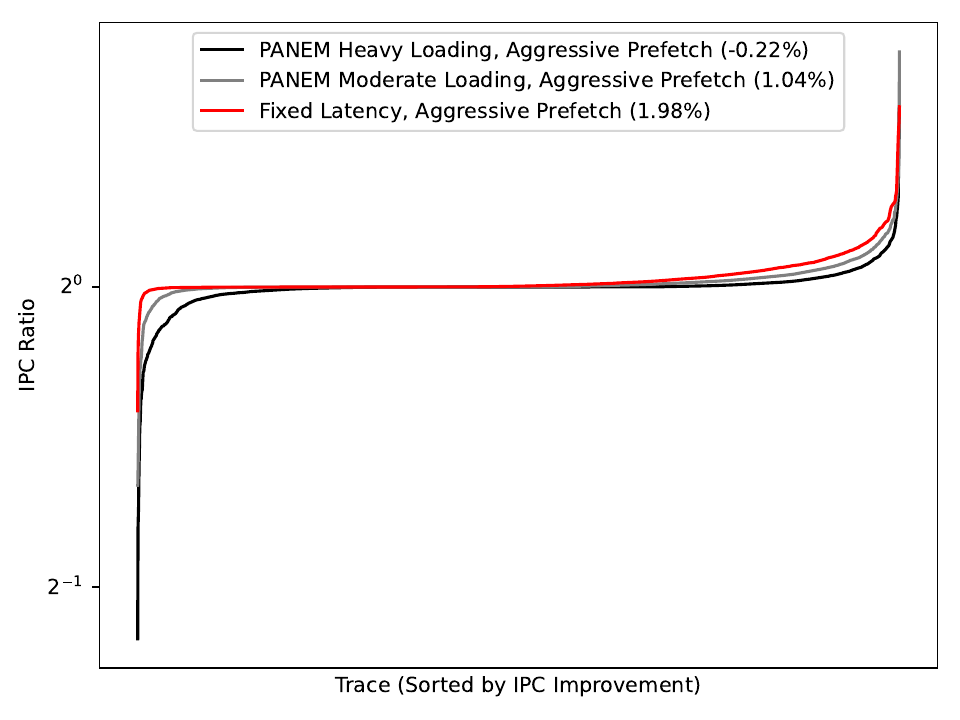}
    \caption{IPC impact of more aggressive L2 prefetching. The heavily loaded PANEM configuration regresses with more prefetches. Cores in a highly loaded system must compete with each other for resources and inaccurate prefetches are a detriment to the entire system's performance.}
    \label{fig:aggressive-prefetch-ucurve}
\end{figure}

 This effect is even more pronounced when increasing the prefetcher aggressiveness (\autoref{fig:aggressive-prefetch-ucurve}). In this case both the PANEM moderate configuration and the fixed-latency model show improvement, but the fixed-latency model is optimistic by a factor of 1.9. PANEM heavy actually slows down; in this case the value of better coverage is not worth the cost in bandwidth utilization, a stark contrast to the prediction of the fixed-latency model.

\subsection{Voluntary Core Throttling}

\begin{figure}[ht] 
    \centering
    \includegraphics[width=\columnwidth]{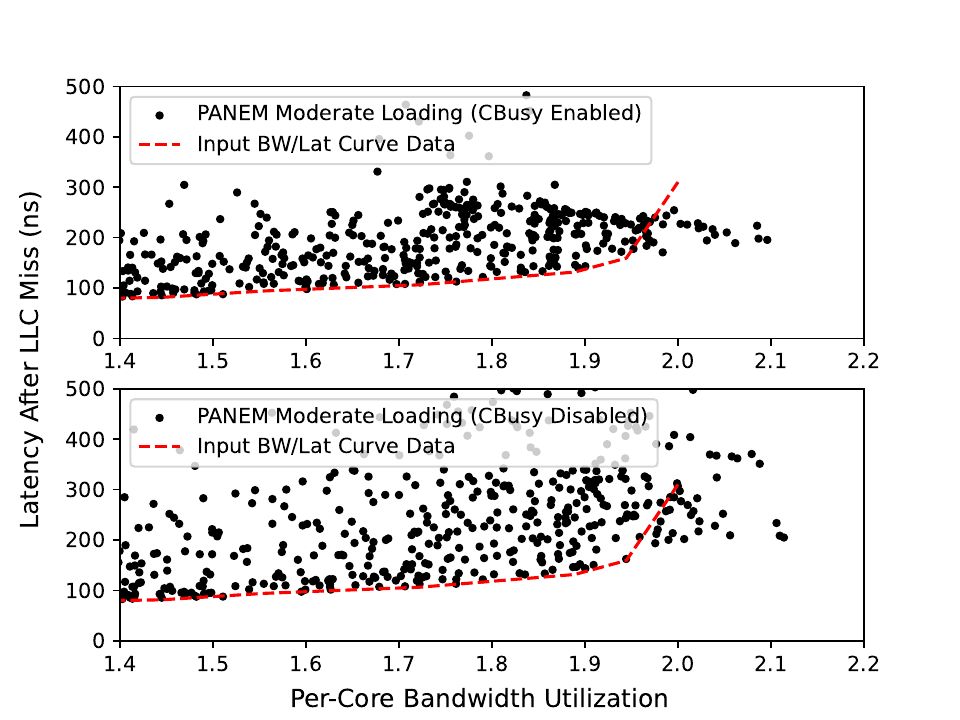}
    \caption{Comparison of workload response to PANEM bandwidth constraints with and without CBusy throttling.}
    \label{fig:panem_cbusy_comparison}
\end{figure}

PANEM also allows us to validate core features intended to improve fair access to system resources within the single-core model. Although insufficient for fine-tuning this behavior, PANEM can evaluate the general efficacy of features like dynamic throttling in response to system load. \autoref{fig:panem_cbusy_comparison} compares CBusy throttling enabled and disabled at the core. Enabling this feature results in a 10\% reduction in average latency across the study list in exchange for a 1.2\% reduction in average bandwidth utilization, thus recovering the latency increase from enabling prefetchers while retaining nearly all of the bandwidth improvement. In practice, this is a pessimistic estimation, especially considering utilization measurement and throttling occur only between core and memory controller in this arrangement, making the control loop delay significantly longer than in a real SoC. The specific implementation of this feature depends on detecting increasing latency from the memory subsystem at higher request rates, making it impossible to reasonably compare PANEM to a fixed-latency model wherein response time is entirely independent of bandwidth demand. Furthermore, fixed-latency models are likely to show a regression under the implementation of any form of throttling as latency cannot be improved and bandwidth is generally unconstrained.

\begin{figure}[ht] 
    \centering
    \includegraphics[width=\columnwidth]{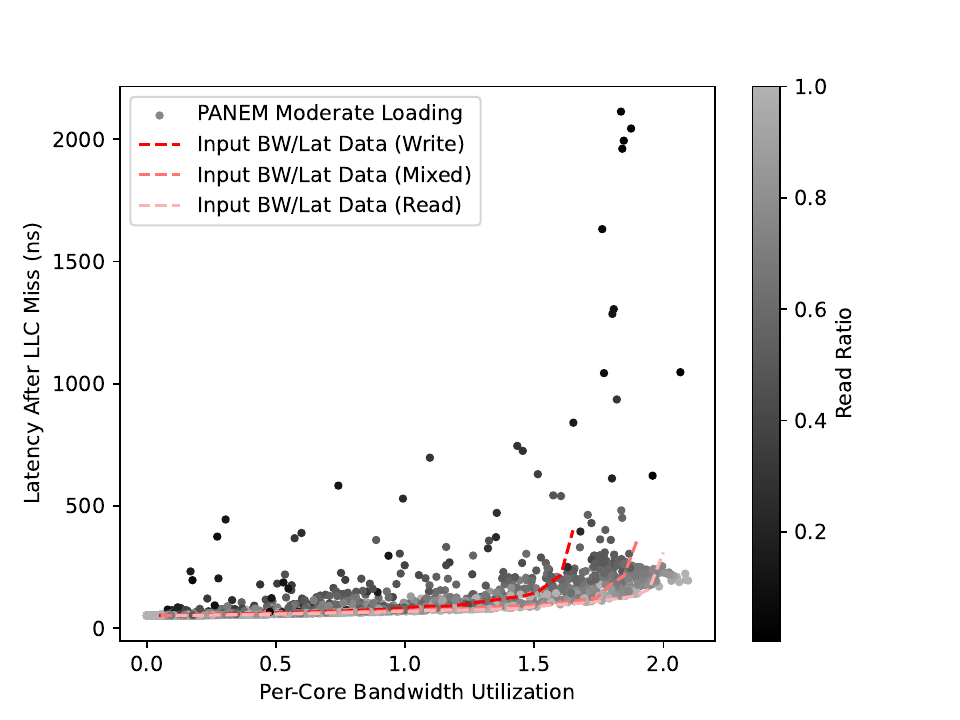}
    \caption{Impact of read ratio on bandwidth and latency. PANEM largely replicates the reported hardware relationship between bandwidth and latency across a wide range of read and write scenarios.}
    \label{fig:read_write_breakdown}
\end{figure}

\autoref{fig:read_write_breakdown} depicts this same configuration with read ratio and corresponding input curve data added. Write-heavy traces tend towards lower bandwidth and higher latency. This is consistent with the input curve definitions and with behavior reported from real systems \cite{mess}. Exceptionally high latencies are limited to write-dominated traces. This is a predictable side-effect of the combination of early write acknowledgment which allows a single core to generate a disproportionately high amount of write bandwidth compared to read bandwidth, the overprovisioned mesh implementation allowing a high degree of write buffering, and the PANEM memory which is scaled down to single- or low-core bandwidth levels. The combined effect of these factors is that latency which would appear distributed throughout the mesh in a real system is lumped into the PANEM processing time for writes.

\subsection{Core Load Depth Sensitivity}
\begin{figure}[ht] 
    \centering
    \includegraphics[width=\columnwidth]{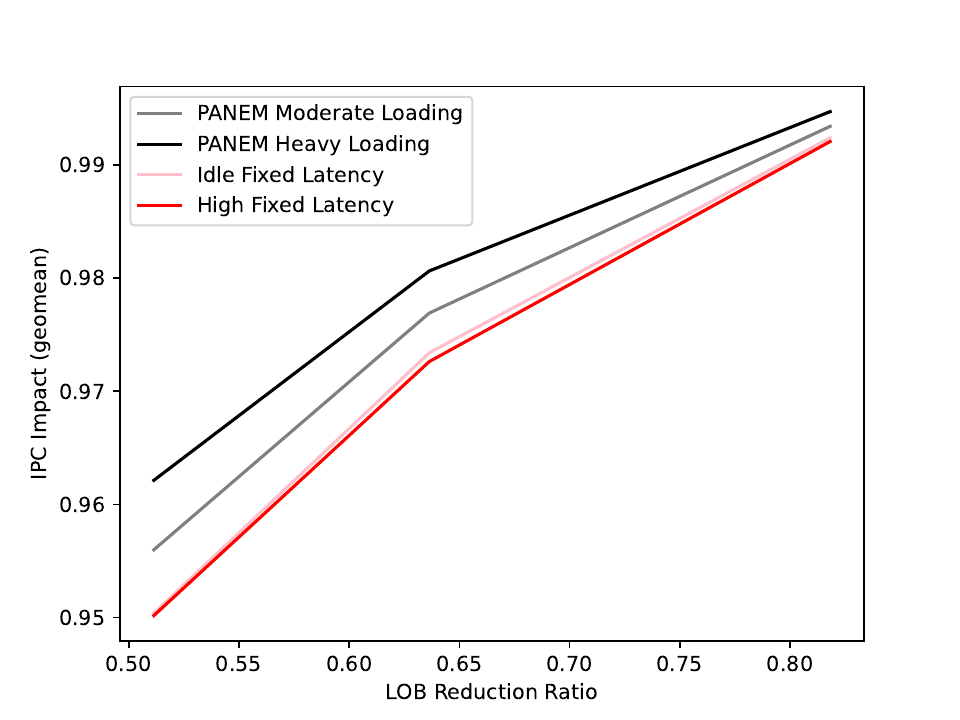}
    \caption{Impact of reducing available load depth. High and low fixed-latency models hardly differentiate while PANEM models expose impact of bandwidth availability.}
    \label{fig:lob_sizing_impact}
\end{figure}

Improvements in predictive capability also extend further into the core. The results of an example Load Ordering Buffer (LOB) size reduction sweep are presented in \autoref{fig:lob_sizing_impact}. The fixed-latency configurations show little differentiation from one another and both suggest a greater sensitivity to LOB size than either PANEM configuration. To incur no more than a 3\% overall IPC impact, idle fixed latency indicates a 3\% larger LOB than PANEM moderate while high fixed latency calls for a nearly 10\% larger LOB than PANEM heavy. Relying on such models while trying to evaluate core resource allocation could result in significant impacts on the efficient deployment of power, area, and timing budgets toward core features.

\subsection{Runtime Impact}
\begin{table}[!h]
    \centering
    \begin{tabular}{c|ccc}
      \textbf{Utilization}  & \textbf{PANEM Runtime}  & \textbf{Fixed Runtime}  &  \textbf{Impact} \\
      \hline
      Moderate & 1913.7 & 1868.3 & 2.43\% \\
      Heavy & 1931.6 & 1911.1 & 1.07\% \\
    \end{tabular}
    \caption{Runtime impact (averaged over 4300 study list data points) of using PANEM compared to fixed-latency models in scheduled CPU seconds.}
    \label{tab:panem_runtime_impact}
\end{table}

\begin{table}[!h]
    \centering
    \begin{tabular}{c|ccc}
      \textbf{Utilization}  & \textbf{PANEM Cycles}  &  \textbf{Fixed Cycles} & \textbf{Relative Cycles/Sec} \\
      \hline
      Moderate & 7279458 & 5875632 & 20.95\% \\
      Heavy & 9825784 & 8113624 & 19.82\% \\
    \end{tabular}
    \caption{Modeled cycles (averaged over 4300 study list data points) for PANEM-based runs and relative efficiency of model.}
    \label{tab:panem_runtime_efficiency}
\end{table}

\begin{figure}[ht] 
    \centering
    \includegraphics[width=\columnwidth]{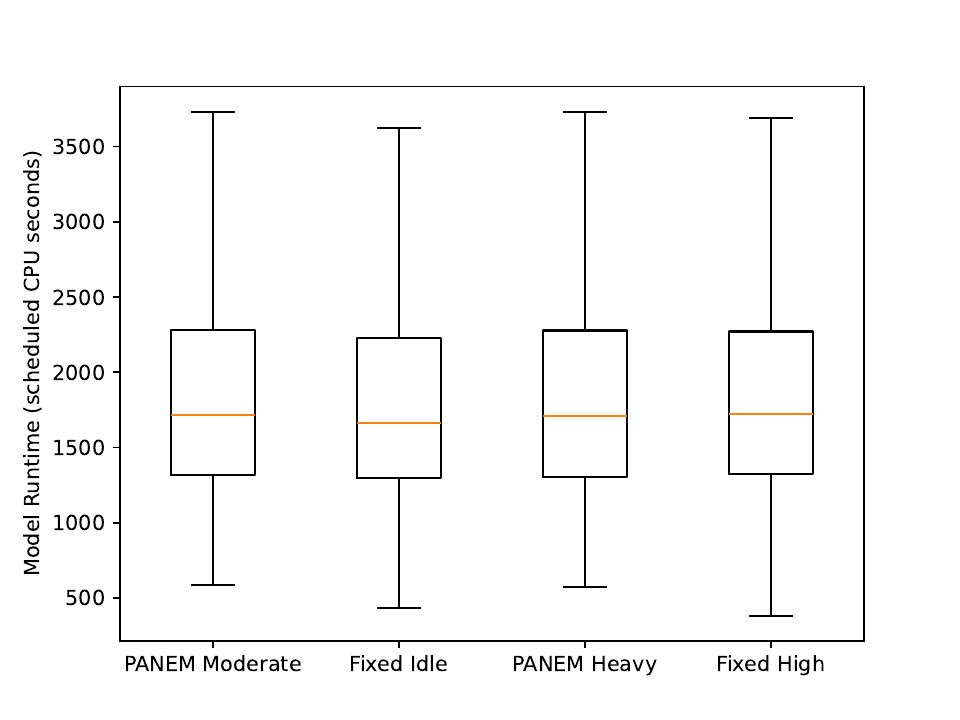}
    \caption{Runtime for the compared configurations.}
    \label{fig:panem_runtime_boxplot}
\end{figure}

The benefits of improved accuracy can only be realized if they are achievable without significantly reducing study throughput. Compared to the fixed-latency models we assessed PANEM against, total mean model runtime impact is minimal and is shown in \autoref{tab:panem_runtime_impact}. When considering the increase in modeled cycles caused by higher latency in PANEM the small impact is even more pronounced: PANEM-based runs model 20\% more cycles per scheduled CPU second than fixed-latency runs (\autoref{tab:panem_runtime_efficiency}). \autoref{fig:panem_runtime_boxplot} summarizes overall model runtime for the four configurations as box plots.

%% file: 60-limitations.tex
\section{Limitations} \label{sec:limitations}
One caveat of PANEM's request-bytes-based measuring scheme is that a sufficient amount of the idle latency within the request path must necessarily be lumped within the curves that parameterize PANEM's behavior. The model is dependent upon new requests seeing old requests still outstanding in order to increase response latency. If the response time of an initial request is too short, then it can respond before an adjacent request is received. This second request will therefore also see the idle latency and it is possible that a sufficient number of simultaneous outstanding requests is never observed to achieve an increase in latency. If accuracy is sought by adding more and more modeled components in front of the PANEM abstraction, less and less of the idle latency is lumped in the curve and this effect can become a significant source of error. Additionally, for single-core simulations where the mesh model connecting core to memory is disproportionate, PANEM will necessarily subsume the fraction of overall latency impact that would otherwise be distributed among the mesh. To account for this, the input curve for PANEM could be adjusted to represent only memory latency at idle request rates with an increasing fraction of the unmodeled mesh latency included at higher request rates. However, in practice, realistic curves tend to have a sharp increase in latency around the bandwidth peak and so the benefit of such an adjustment would be minimal: latency will increase dramatically if requests exceed capacity and the onset of this behavior will occur at nearly identical request rates regardless of whether or not this effect is accounted for.
\begin{figure}[ht] 
    \centering
    \includegraphics[width=\columnwidth]{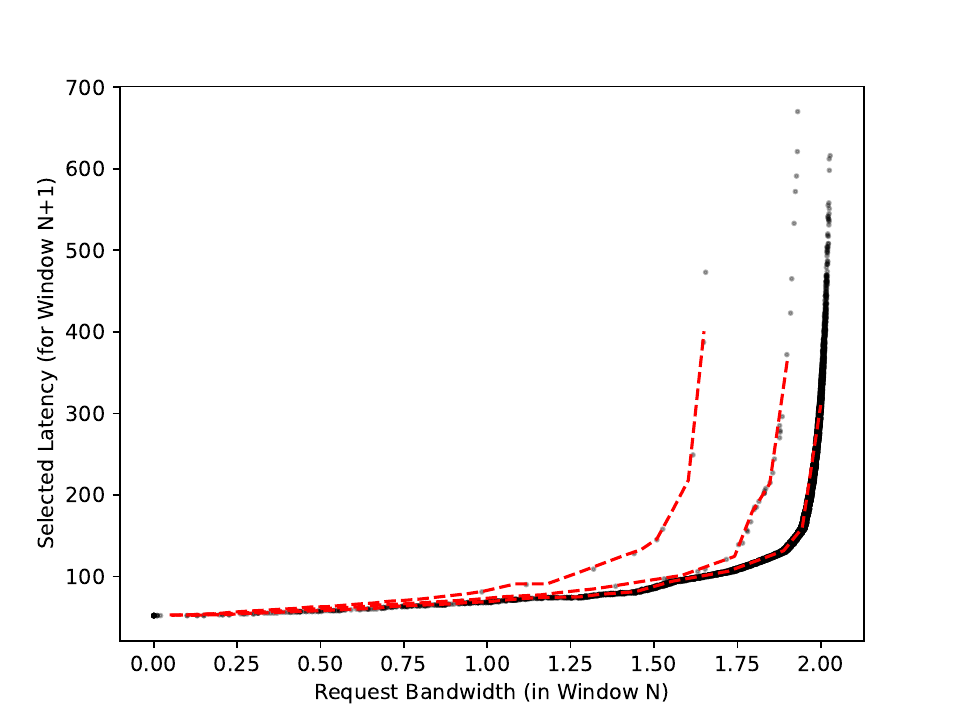}
    \caption{Request bandwidth in each window and resulting latency applied to the following window within highest bandwidth trace. Latency closely follows input curves. Unlike delivered bandwidth, request rate is not directly impacted by latency and can continue to increase above the rate at which peak delivered bandwidth is achieved.}
    \label{fig:requests_vs_latency_selection}
\end{figure}

\begin{figure}[h] 
    \centering
    \includegraphics[width=\columnwidth]{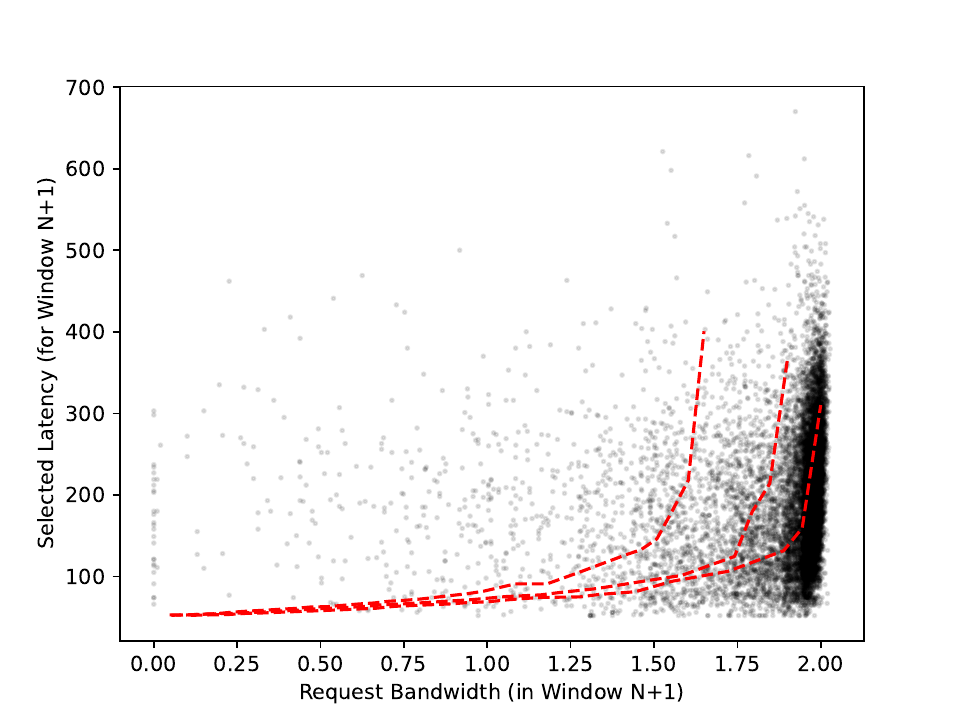}
    \caption{Request bandwidth in each window and previously selected latency applied to that window within highest bandwidth trace. A large number of high bandwidth windows follow windows in which a low latency is selected.}
    \label{fig:requests_vs_effective_latency}
\end{figure}

\begin{figure}[ht] 
    \centering
    \includegraphics[width=\columnwidth]{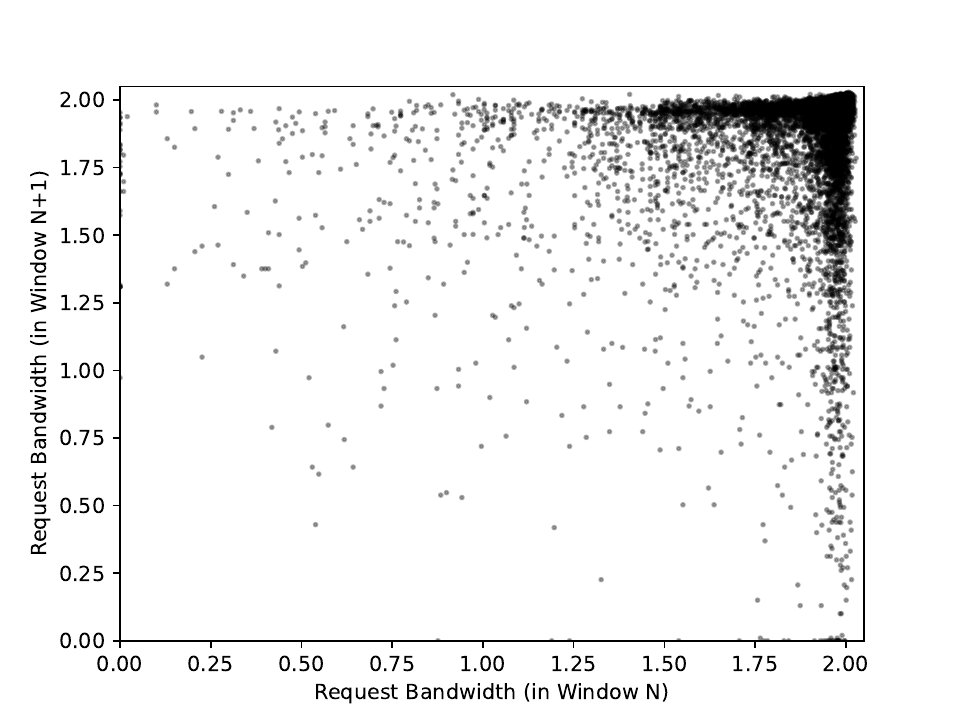}
    \caption{Request bandwidth in adjacent pairs of windows within highest bandwidth trace. The moderate amount of anticorrelation leads to alternating selection of low and high latencies.}
    \label{fig:requests_autocorrelation}
\end{figure}

\begin{figure}[ht] 
    \centering
    \includegraphics[width=\columnwidth]{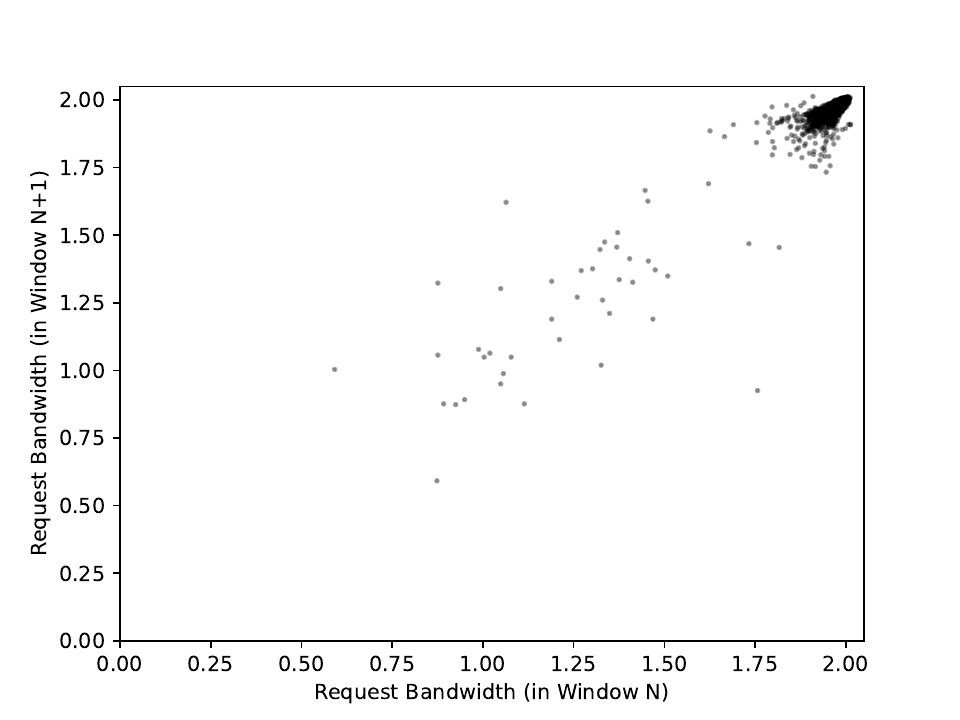}
    \caption{Request bandwidth in adjacent pairs of windows within highest bandwidth trace. Increasing averaging window to $4 \times 520$ ns results in a high degree of correlation between bandwidth in adjacent windows.}
    \label{fig:requests_autocorrelation_wider}
\end{figure}

Although the vast majority of traces are bound by the supplied bandwidth/latency curve or its extrapolation via the computed request-bytes/latency curve, some traces do manage to achieve a latency and bandwidth combination outside of the curve bounds. The traces which outright exceed the peak bandwidth account for 0.32\% of traces and the highest bandwidth trace achieves an average bandwidth 4.8\% above the theoretical peak. Traces achieving a latency lower than expected for their given average bandwidth regardless of whether the peak bandwidth is exceeded account for 0.57\% of traces. All of these traces tend to exhibit a bipolar distribution of high and moderate requested bandwidth with a period similar to that of the chosen averaging window size. Because latency selection follows request measurement, a situation can arise where high latencies are applied to adjacent lower bandwidth request regions and lower latencies are applied to adjacent higher bandwidth request regions. The net result is that a trace which consistently consumes bandwidth in one window but not in the next window will achieve a high average bandwidth at a low average latency (highest bandwidth trace depicted in Figures \ref{fig:requests_vs_latency_selection}, \ref{fig:requests_vs_effective_latency}, and \ref{fig:requests_autocorrelation}).

\begin{figure}[hb] 
    \centering
    \includegraphics[width=\columnwidth]{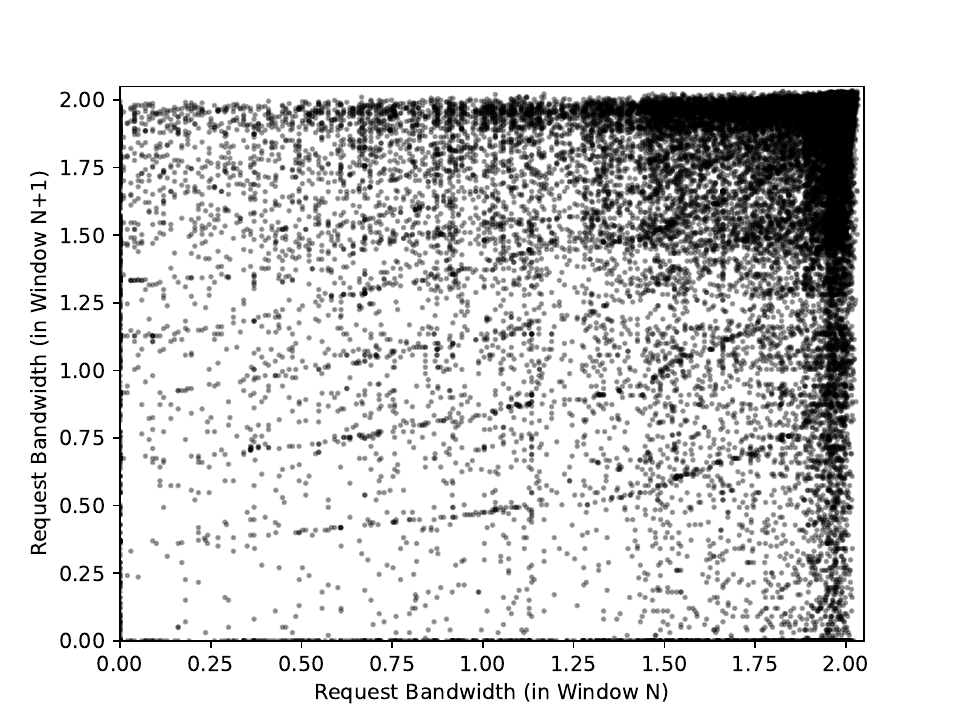}
    \caption{Request bandwidth in adjacent pairs of windows within highest bandwidth trace. Reducing averaging window to $1 \times 210$ ns exposes detail of workload request behavior. Bandwidth is also limited in this case but is driven by rapid reaction to changes in request rate, not sustained latency.}
    \label{fig:requests_autocorrelation_narrower}
\end{figure}

\begin{table}
    \centering
    \begin{tabular}{c|ccc}
      \textbf{Window}  & \textbf{Normalized BW}  & \textbf{Latency (ns)}  &  \textbf{Relative IPC}\\
      \hline
      $1 x 210 $ ns & 1.99 & 162 & 0.954 \\
      \textit{$1 x 520 $ ns} & \textit{2.09} & \textit{196} & \textit{1.000} \\
      $2 x 520 $ ns & 2.07 & 213 & 0.987 \\
      $4 x 520 $ ns & 2.01 & 227 & 0.957 \\
    \end{tabular}
    \caption{Impact of window size selection on highest bandwidth trace.}
    \label{tab:window_size_impact}
\end{table}

Choosing a different window size can cause these traces to no longer exceed the peak bandwidth. For single-core simulations we generally desire slight homogenization of bandwidth as achieved with the wider averaging window used to generate \autoref{fig:requests_autocorrelation_wider}. This helps ensure PANEM is reacting as if it were seeing a field of many cores making similar but orthogonal requests as that of the simulated core, rather than reacting to the particular behavior of the single simulated core. Nevertheless, choosing a smaller window size like that used to generate \autoref{fig:requests_autocorrelation_narrower} also results in a reduction in bandwidth and at a lower latency. In this case, specific characteristics of the workload executing on the single modeled core are reflected in successive windows. These effects are summarized in \autoref{tab:window_size_impact} and suggest that the default window size happens to be finely tuned to the behavior of this trace which allows it to consume bandwidth without paying a fully appropriate penalty. We have observed that for any reasonable selection of window size, there are corresponding traces exhibiting periodic request intensity which will cause them to similarly exceed the bandwidth peak. Thus, there is not a single window size equally appropriate to all traces within our study list. Adapting window size to individual workload behavior may be an effective method for improving accuracy, but given the very low incidence of non-compliant traces and the relatively low deviation in even the highest bandwidth trace, the opportunity seems limited. This effect motivated our default window size selection; we chose the smallest window that had no individual trace-average outliers greater than 5\% of nominal peak bandwidth. We believe that this choice maximizes workload reactivity while mitigating impacts from unintended bandwidth capacity.

%% file: 80-conclusion.tex
\section{Conclusion}

This work presented PANEM, a lightweight event-driven heuristic memory model for single-core simulation under realistic contention, as seen in many-core cloud systems. By deriving a request-bytes/latency response from bandwidth/latency characterization data, PANEM captures dynamic congestion effects while retaining the speed and configurability needed for large pre-silicon study campaigns. Across our workloads, PANEM largely enforced appropriate bandwidth limits for a variety of read/write mixes and reproduced the expected increase in latency at higher demand, unlike fixed-latency models that either overestimate achievable performance or understate available throughput depending on the chosen constant latency point.


The results show that model choice materially changes architectural decisions. Fixed-latency assumptions mispredicted IPC trends (e.g., +6.74\% optimistic in the moderately loaded regime and -3.85\% pessimistic in the heavily loaded regime), distorted sensitivity to prefetch behavior, and biased core resource sizing (including LOB provisioning by roughly 3-10\%). PANEM also enabled meaningful evaluation of control features such as dynamic throttling, demonstrating a practical latency-bandwidth tradeoff under load. Although PANEM is weakly susceptible to interactions between averaging window size and workload periodicity, overall PANEM provides a pragmatic middle ground between simplistic fixed-latency abstractions and expensive cycle-accurate DRAM simulation, improving predictive confidence in high-throughput studies for core and system co-design in commercial cloud SoCs.


%% file: 99-appendix.tex
\begin{figure*}[p]
    \centering
    \includegraphics[
        width=\textwidth,
        height=.9\textheight,
        keepaspectratio
    ]{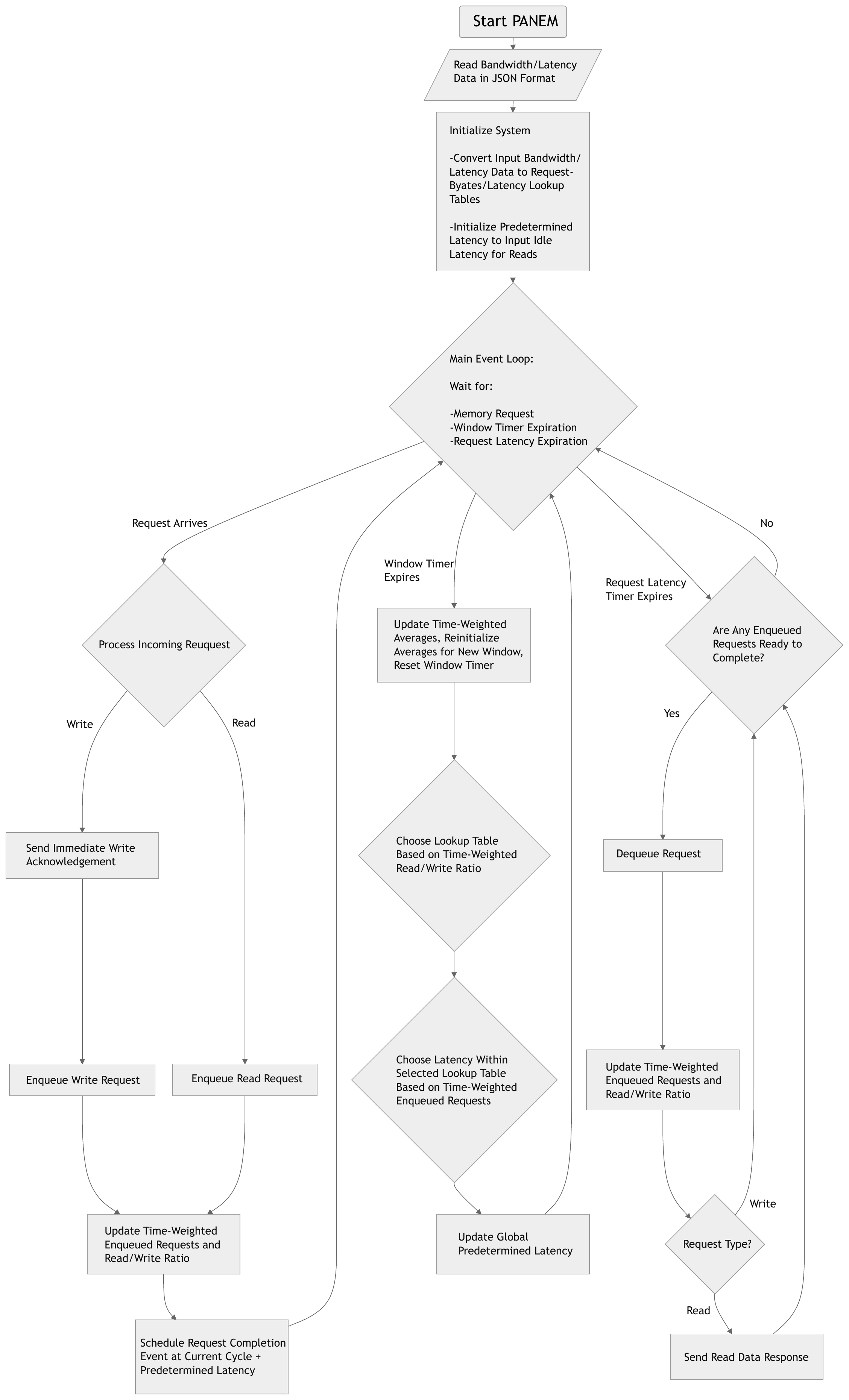}
    \caption{General flow of PANEM event loop. All requests are enqueued for the current latency at the time of their arrival at PANEM. Requests are dequeued when their latency expires and a response and/or data packet may be sent at that time. Writes can early acknowledge so they send a response immediately but still enqueue for the current latency. Whenever a request is enqueued or dequeued the time-weighted average number of outstanding request-bytes is updated. When the latency window timer expires the current time-weighted average outstanding request-bytes is used to lookup a new latency; the outstanding request-bytes average is then initialized to the current outstanding request-bytes value to begin the next averaging window. }
    \label{fig:flowchart}
\end{figure*}